\documentclass[aps,prb,superscriptaddress,twocolumn,showpacs,preprintnumbers,amsmath,amssymb]{revtex4}
\usepackage{color}
\usepackage{array}
\usepackage{amsmath}
\usepackage{amssymb}
\usepackage{epsfig}
\usepackage{graphicx}
\usepackage{wasysym}
\usepackage{bm}

\newcommand{\be}{\begin{equation}}
\newcommand{\ee}{\end{equation}}
\newcommand{\bea}{\begin{eqnarray}}
\newcommand{\eea}{\end{eqnarray}}

\begin{document}

\title{Topological Entanglement Entropy of $Z_2$ Spin liquids and
Lattice Laughlin states}
\author{Yi Zhang, Tarun Grover and Ashvin Vishwanath}

\affiliation{Department of Physics, University of California,
Berkeley, CA 94720}

\begin{abstract}
We study entanglement properties of candidate wave-functions for
$SU(2)$ symmetric gapped spin liquids and Laughlin states. These
wave-functions are obtained by the Gutzwiller projection technique. Using Topological Entanglement Entropy
$\gamma$ as a tool, we establish topological order in chiral spin liquid and
$Z_{2}$ spin liquid wave-functions, as well as a lattice version of the Laughlin
state. Our results agree very well with
the field theoretic result $\gamma=\log D$ where $D$ is the total
quantum dimension of the phase. All calculations are done using a
Monte Carlo technique on a $12 \times 12$ lattice enabling us to
extract $\gamma$ with small finite size effects. For a chiral spin
liquid wave-function, the calculated value is within $4\%$ of the
ideal value. We also find good agreement for a lattice version of
the Laughlin $\nu=1/3$ phase with the expected
$\gamma=\log\sqrt{3}$.
\end{abstract}

\maketitle

\section{Introduction}

Quantum Spin Liquids(SLs) are states that arise from the collective
behavior of spins, but are not characterized by a Landau order
parameter. They are associated with remarkable phenomena such as
fractional quantum numbers \cite{anderson1987}, transmutation of
statistics (eg. fermions appearing in a purely bosonic
model)\cite{kivelson1987, read1989}, and enabling otherwise
impossible quantum phase transitions\cite{senthil2004}, to name a
few. SLs may be gapless or gapped. While current experimental
candidates for SLs appear to have gapless excitations \cite{exp},
gapped SLs are indicated in numerical studies on the Kagome
\cite{yan2010} and honeycomb lattice \cite{meng2010}. Gapped SLs are
characterized by topological order - i.e. ground state degeneracy
that depends on the topology of the underlying space\cite{wen2004}.

Recently, a novel characterization of gapped SLs has emerged using
quantum entanglement in terms of the Topological Entanglement
Entropy (TEE)\cite{hamma2005,levin2006, kitaev2006}. This quantity
takes a fixed value $\gamma$ in a topologically ordered phase and
remarkably can be calculated just knowing the ground state
wave-function. The entanglement entropy of a two dimensional disc
shaped region A in a gapped phase obeys $S_{2}=al_{A}-\gamma$, where
a smooth boundary of length $l_{A}$ is assumed to surround the
region. By carefully subtracting off the leading dependence, the
constant $\gamma$ can be isolated. It is argued to be a
characteristic of the phase, $\gamma=\log D$, where $D$ is the
quantum dimension of the phase \cite{levin2006, kitaev2006}. For the
abelian states discussed here, $D^2$ is identical to the ground
state degeneracy on the torus.

Gapped SLs can be viewed as a state where each spin forms a singlet
with a near neighbor, but the arrangement of singlets fluctuates
quantum mechanically so it is a liquid of singlets. Theoretical
models of this singlet liquid fall roughly into two categories. In
the first, the singlets are represented as microscopic variables as
in quantum dimer and related models\cite{rokhsar1988, moessner2001,
senthil2002, balents2002,misguich}, and are suggested by large N
calculations \cite{sachdev1992}. Topological order can then be
established by a variety of techniques including exact solution and
most recently quantum entanglement \cite{Furukawa, Claudio, Fradkin,
isakov2011,hamma2008}. In contrast there has been less progress
establishing topological order in the second category, which are
SU(2) symmetric spin systems where valence bonds are emergent
degrees of freedom. Anderson\cite{anderson1987} proposed
constructing an SU(2) symmetric SL wave-function by starting with a
BCS state, derived from the mean field Hamiltonian: \be H =
-\sum_{rr'} \{t_{rr'}f^{\dagger}_{\sigma,r} f_{\sigma,r'} +
\Delta_{r r'} f^{\dagger}_{\uparrow,r} f^{\dagger}_{\downarrow,r'}\}
+ h.c. \ee and Gutzwiller projecting it so that there is exactly one
fermion per site, hence a spin wave-function. Variants of these are
known to be good variational ground states for local spin
Hamiltonians (see e.g. \cite{varstudy}) and are more viable
descriptions of most experimental and $SU(2)$ symmetric liquids.
Approximate analytical treatments of projection, that include small
fluctuations about the above mean field state, indicate that at
least two kinds of gapped SLs can arise: chiral SLs\cite{kalmeyer}
and $Z_2$ SLs \cite{sachdev1992,wen1991,senthil2000}. However, given
the drastic nature of projection, it is unclear if the actual
wave-functions obtained from this procedure are in the same phases.
In this paper we use TEE to establish topological order for of
$SU(2)$ symmetric chiral and $Z_2$ SL wave-functions. We show that
the recently developed Monte Carlo technique used to study
entanglement properties of gapless SLs\cite{frank2011} can be
applied here as well to extract TEE for system sizes large enough
(144 spins) so that it approaches its quantized value. Instead of
using the more standard Von Neumann entropy, we focus on the Renyi
entropy, which carries the same contribution from the TEE for both
 non-chiral \cite{flammia2009} and chiral \cite{dong} states for a topologically trivial bipartition. The fact that
the wave-function is a determinant or product of determinants in
these cases allows for its efficient evaluation. For a model of of
the chiral SL, the calculated TEE are remarkably accurate, within
few percent of the expected $\log \sqrt{2}$ value. To our knowledge,
this is the first clear demonstration of topological order via TEE,
in SU(2) symmetric spin wave-functions.

We also study lattice versions of the Laughlin $\nu=1/3$ state,
which are obtained by a similar projective construction, although
these are fermionic, not spin wave-functions. Again we can extract
TEE which is within 7\% of the expected value to confirm these are
in the same phase as the Laughlin state, although they differ
significantly in microscopic structure.

We note earlier numerical work extracting TEE include exact
digitalization studies on small systems, looking at quantum Hall
Laughlin states \cite{haque2007} and perturbed Kitaev toric code
models\cite{hamma2008}. Recently, a quantum Monte Carlo study
\cite{isakov2011} used TEE to detect $Z_{2}$ topological order. In
contrast to the states studied here, this was a positive definite
wave-function, with U(1) rather than SU(2) spin symmetry. Our
wave-function-only approach is ready-made for searching for
topological order when one has a good variational ansatz for a
ground state, irrespective of whether the state is positive definite
or not. Finally, we note that Ref. \cite{ivanov2002} studied
topological order in `nodal' $Z_2$ SLs by constructing four
orthogonal low energy states on the torus, and Ref. \cite{Yao}
studied TEE for the Kitaev model.

The format of the paper and main results are summarized in the
section below.

In Section \ref{sectionTEE} the TEE $\gamma$ is defined, and an
algorithm to calculate it numerically utilizing the Renyi
entanglement entropy $S_2$ is outlined. This is then applied to a
series of topological phases, the results of which are summarized in
table \ref{table1}.

(i) The first is a Chern insulator, built out of a square lattice
tight binding model at half filling, in which the filled band has
unit Chern number. For a lattice with $2N$ sites, this is an N body
Slater determinant $\Phi(r_1,\dots, \,r_N)$. Since this is an
integer quantum Hall state, it is not expected to possess
topological order. Indeed, calculation is consistent with a
vanishing TEE, see Table \ref{table1} first row.

(ii) The chiral SL wave-function is obtained from the wave-function
of 2N spinful electrons with this tight binding band structure, by
projecting out all double occupancies, and studied in Section
\ref{sectionChiralSL}. The chiral SL wave-function can be written as
a product of two Slater determinants,
i.e.$\Psi(r_1,\,r_2\dots,\,r_N)={\mathcal M}\Phi^2(r_1,\dots,
\,r_N)$,  where ${\mathcal M}$ an unimportant Marshall sign factor.
If we view up spin as a hardcore boson, then this is the
wave-function analogous to half filled Landau level $\nu=1/2$ of
bosons. It is therefore expected to have $\gamma=\log \sqrt{2}$.
Indeed, for a particular choice of parameters with a large gap,
numerical calculation (second and third rows of table \ref{table1},
with different linear dimensions $L_A$ of the smallest subregions
involved) yields a value very close to this. Detailed finite size
analysis obtained by varying the correlation length of the chiral SL
is presented in Section \ref{sectionChiralSL}, providing further
evidence for convergence to the expected value. We note that this
wave-function is SU(2) symmetric, and non-positive-definite, since
the ground state is not time reversal symmetric.

(iii) Note, the construction of the chiral SL above is similar to
the Laughlin construction of fractional quantum Hall states by
taking products of the integer quantum Hall states. Extending the
construction above, one can write wave-functions for $N$ fermions
$\Psi_{1/3}(r_1,\,r_2\dots,\,r_N)=\Phi^3(r_1,\dots, \,r_N)$, a
lattice version of the Laughlin $\nu=1/3$ state. The entanglement
entropy calculation for $\Psi_{1/3}$ agrees well with what is
expected for the topological order for $\nu=1/3$ Laughlin state,
indicating it is in the same phase, despite not being constructed
from lowest Landau level states. Note, since they differ
significantly in microscopic detail from the Laughlin state,
wave-function overlap is not an option in establishing that they are
in the same phase. Also, calculating entanglement spectra
\cite{li2008} is currently not feasible for these wave-functions,
thus TEE appears to be the ideal characterization. Similarly, the
lattice analog of Laughlin $\nu=1/4$ state for bosons, obtained via
$\Psi_{1/4}(r_1,\,r_2\dots,\,r_N)=\Phi^4(r_1,\dots, \,r_N)$, is
found to have a TEE close to the expected $\gamma=\log \sqrt{4}$, as
discussed in Section \ref{sectionLaughlin}.

(iv) Finally, we construct a fully gapped $Z_2$ SL wave-function on
the square lattice. For the largest system sizes we considered, the
calculated $\gamma$ is $84\%$ of the expected $\log 2$ value(last
row in table \ref{table1}). The difference is ascribed to larger
finite size effects, as discussed in Section \ref{Z2}.

\begin{table}
\begin{tabular}{|l|l|l|}
  \hline
  State & Expected $\gamma$ & $\gamma_{\rm calculated}/\gamma_{\rm expected}$ \\
    \hline
  Unprojected ($\nu=1$) & 0 &  -0.0008$\pm$ 0.0059 $^{*}$ \\
  Chiral SL L$_A$=3  & $\log\sqrt{2}$ & 0.99 $\pm$  0.03\\
  Chiral SL L$_A$=4  & $\log\sqrt{2}$ & 0.99 $\pm$ 0.12 \\
  Lattice $\nu=1/3$   & $\log\sqrt{3}$ & 1.07$\pm$ 0.05 \\
  Lattice $\nu=1/4$    & $\log\sqrt{4}$ & 1.06 $\pm$ 0.11\\
  Z$_2$ SL L$_A$=4 & $\log{2}$ & 0.84 $\pm$ 0.13\\ \hline
\end{tabular}
\caption{Comparison between calculated TEE and expected value from
field theory(second column) for topological phases. The $^*$ denotes
that the calculated value is not divided by the expected value since
the latter vanishes.} \label{table1}
\end{table}

\section{Topological entanglement entropy and Variational Monte Carlo method}
\label{sectionTEE}
\subsection{Renyi entropy and topological entanglement entropy}

Given a normalized wave-function $\left|\Phi\right\rangle$ and a
partition of the system into subsystems $A$ and $B$, one can trace
out the subsystem $B$ to obtain the reduced density matrix on $A$:
$\rho_{A}=Tr_{B}\left|\Phi\right\rangle\left\langle\Phi\right|$. The
Renyi entropies are defined as:
\begin{equation}
S_{n}=\frac{1}{1-n}\log\left(Tr\rho^{n}_{A}\right) \label{Renyi}
\end{equation}

Taking the limit $n\rightarrow 1$, this recovers the definition of
the usual von Neumann entropy. In this paper we will focus on the
Renyi entropy with index $n=2$:
$S_{2}=-\log\left(Tr\left(\rho^{2}_{A}\right)\right)$, which is
easier to calculate with our Variational Monte Carlo(VMC) method
\cite{frank2011}.

For a gapped phase in 2D with topological order, a contractible
region $A$ with smooth boundary of length $l_{A}$, the Area Law of
the Renyi entropy becomes:
\begin{eqnarray*}
S_{2}=al_{A}-\gamma
\label{eqntee0}
\end{eqnarray*}

where we have omitted the sub-leading terms. Although the
coefficient $a$ of the leading boundary law' term is non-universal,
the sub-leading constant $\gamma$ is universal, and this TEE is a
robust property of the phase of matter for which $|\Phi\rangle$ is
the ground state. It is given by $\gamma=\log D$, where $D$ is the
total quantum dimension of the model \cite{levin2006, kitaev2006},
and offers a partial characterization of the underlying topological
order. When region $A$ has a disc geometry, it has been shown that
$\gamma$ for different Renyi indices $n$ are identical for both
chiral and non-chiral states\cite{flammia2009, dong}. A simple limit
where this is readily observed\cite{hamma2005,levin2006} is in a
model wave-function of a $Z_2$ SL, which is an equal superposition
of loops ($Z_2$ electric field). This is achieved as a ground state
in Kitaev's toric code model \cite{kitaev2003}. The Schmidt
decomposition into wave-functions in regions $A$ and $B$ can be
indexed by the configuration of electric field lines piercing the
boundary of the disc. If  $i=1,2,\dots l$ are $l$ bonds going
through the boundary between region $A$ and $B$, the presence
(absence) of electric field lines on bond $i$ is denoted by $q_i=1$
($q_i=0$). Since the loops are closed, we require $\sum_iq_i={\rm
even}$. It can be shown that the wave-function is simply an equal
weight decomposition indexed by all possible configurations of
$q_i$. There are $C=2^{l-1}$ of them, the global constraint of
closed loops accounting for the missing factor of $2$. Then:
\begin{equation*}|\Psi\rangle=\frac1{\sqrt{C}}\sum_{q_1+\dots+q_l\,{\rm
even}}|\Psi^A_{q_1\dots q_l}\rangle|\Psi^B_{q_1\dots q_l}\rangle\end{equation*}

This implies\cite{levin2006} there are $C$ equal eigenvalues of the
region A density matrix, each equal to $1/C$. The Renyi entropy from
Eqn.\ref{Renyi} is: $S_n=\frac1{1-n}\log C^{-(n-1)}=(l-1)\log 2$.
Thus $\gamma=\log 2$ from the definition above, if we identify $l$
with the length of the boundary. Note, this follows independent of
the Renyi index $n$, and is the expected value for a $Z_2$ gauge
theory with quantum dimension $D=2$.

Practically, it is not convenient to extract the subleading constant
by fitting the expression above, particularly on the lattice where
edges frequently occur. Instead, one may use a construction due to
Levin and Wen\cite{levin2006}, or Kitaev and
Preskill\cite{kitaev2006}, that effectively cancels out the leading
term and exposes the topological contribution. We use the latter,
which requires calculating entanglement entropy for a triad of non
overlapping regions A, B, C, and their various unions, and then
constructing:

\begin{eqnarray}
-\gamma&=&S_{A}+S_{B}+S_{C}-S_{AB}-S_{AC}-S_{BC}+S_{ABC} \nonumber
\label{gamma}
\end{eqnarray}

here, any $S_n$ can be used, and we choose to use $S_2$, since it
can be easily calculated. This guarantees that the contributions of
boundaries and corners cancel when the dimensions of individual
regions $A, B,C$ is much larger than the correlation length.

\subsection{Variational Monte Carlo method for Renyi Entropy}
In this section we briefly review the VMC algorithm for calculating
Renyi entropy $S_2$ \cite{hastings2010, frank2011}. Consider the
configurations $|\alpha_{1}\rangle =|a\rangle |b\rangle$, $
|\alpha_{2}\rangle = |a'\rangle |b'\rangle $, $|\beta_{1}\rangle
=|a'\rangle |b\rangle $, $|\beta_{2}\rangle = |a\rangle |b'\rangle$,
where $|a\rangle$ and $|a'\rangle$ have their support only in the
subsystem $A$ while $|b\rangle$ and $|b'\rangle$ are in subsystem
$B$. Following Ref.\cite{hastings2010}, we define an operator
$\rm{Swap_{A}}$ that acts on the tensor product of two copies of the
system and swaps the configurations of the spins belonging to the
$A$ subsystem in the two copies i.e. $\rm{Swap_{A}}
|\alpha_{1}\rangle \otimes |\alpha_2 \rangle = |\beta_{1}\rangle
\otimes |\beta_2 \rangle $. The Renyi entropy $S_2$ for the
bipartition $A$ and $B$ can be expressed in terms of the expectation
value of $\rm{Swap_{A}}$ with respect to the wave-function
$|\Phi\rangle \otimes |\Phi\rangle$:

\begin{eqnarray}
S_{A}
=-\log\left(tr\rho_{A}^{2}\right)=-\log\left<\rm{Swap_{A}}\right>
\label{eqntee1}
\end{eqnarray}

$<\rm{Swap_{A}}>$ may be re-expressed as a Monte Carlo average:

\begin{eqnarray}
\left<\rm{Swap_{A}}\right>=\underset{\alpha_{1},\alpha_{2}}{\sum}\rho_{\alpha_{1}}\rho_{\alpha_{2}}f\left(\alpha_{1},\alpha_{2}\right)
\label{eqntee2}
\end{eqnarray}

where the weights
$\rho_{\alpha_{i}}=\left|\left\langle\alpha_{i}|\Phi\right\rangle\right|^{2}/\sum_{\alpha_{i}}\left|\left\langle\alpha_{i}|\Phi\right\rangle\right|^{2}$
are normalized and non-negative while the quantity to be averaged
over the probability distribution
$\rho_{\alpha_{1}}\rho_{\alpha_{2}}$ is:

\begin{eqnarray}
f\left(\alpha_{1},\alpha_{2}\right)=\frac{ \left\langle\beta_{1}|\Phi\right\rangle \left\langle\beta_{2} |\Phi\right\rangle} { \left\langle\alpha_{1}|\Phi\right\rangle \left\langle\alpha_{2} |\Phi\right\rangle }
\label{eqntee3}
\end{eqnarray}

Therefore, one can calculate the Renyi entropy using VMC method.
This technique is particularly suited for projected wave-functions
since the projection is rather easy to implement in a VMC algorithm
\cite{gros1989}. As shown in Ref. \cite{frank2011} VMC algorithm
correctly reproduces the exact results for free fermions with an
error of less than a few percent.

We further facilitate our calculation with an algorithm that we
referred to as the sign trick\cite{frank2011}. It offers
simplification and reduces computational cost. Basically, we
separate $\left<\rm{Swap_{A}}\right>$ as a product of two factors,
which may be independently calculated within VMC method:

\begin{eqnarray*}
\left\langle \rm{Swap_{A}}\right\rangle & = & \langle
\rm{Swap_{A,mod}}\rangle \langle \rm{Swap_{A,sign}}\rangle \\ &=&
\underset{\alpha_{1}\alpha_{2}}{\sum}\rho_{\alpha_{1}}\rho_{\alpha_{2}}|f\left(\alpha_{1},\alpha_{2}\right)|
\left [
\underset{\alpha_{1}\alpha_{2}}{\sum}\tilde{\rho}_{\alpha_{1},\alpha_2}e^{i\phi
\left(\alpha_{1},\alpha_{2}\right)}\right]
\end{eqnarray*}

The first factor is the Renyi entropy of a sign problem free
wave-function $|\phi_{\alpha_i}|$. The second term is the
expectation value of the phase factor $e^{ i\phi
\left(\alpha_{1},\alpha_{2}\right)} =
\phi^*_{\alpha_1}\phi^*_{\alpha_2}\phi_{\beta_1}\phi_{\beta_2}/\left|\phi^*_{\alpha_1}\phi^*_{\alpha_2}\phi_{\beta_1}\phi_{\beta_2}\right|$
with probability distribution $\tilde{\rho}_{\alpha_{1},\alpha_2} =
|\phi^*_{\alpha_1}\phi^*_{\alpha_2}\phi_{\beta_1}\phi_{\beta_2}|/
\underset{\alpha_{1}\alpha_{2}}{\sum}|\phi^*_{\alpha_1}\phi^*_{\alpha_2}\phi_{\beta_1}\phi_{\beta_2}|$.
Both factors can be calculated in a more efficient manner and most
importantly, have much smaller errors than the direct calculation of
$\left\langle \rm{Swap_{A}}\right\rangle$.

\section{Entanglement entropy for a chiral spin liquid}
\label{sectionChiralSL} In this section we calculate the Renyi
entropy $S_2$ and TEE $\gamma$ for a chiral SL\cite{kalmeyer}.

\subsection{Projected wave-function for chiral spin liquids}

The chiral SL is a spin SU(2) singlet ground state, that breaks time
reversal and parity symmetry\cite{kalmeyer,thomale}. A wave-function
in this phase wave function may be obtained using the slave-particle
formalism by Gutzwiller projecting a $d+id$ BCS state
\cite{kalmeyer}. Alternately, it can be obtained by Gutzwiller
projection of a hopping model on the square lattice. This model has
fermions hopping on the square lattice with a $\pi$ flux through
every plaquette and imaginary hoppings across the square lattice
diagonals:

\begin{eqnarray}
H=\underset{\left\langle ij\right\rangle
}{\sum}t_{ij}f_{i}^{\dagger}f_{j}+ i\underset{\left\langle
\left\langle ik\right\rangle \right\rangle
}{\sum}\Delta_{ik}f_{i}^{\dagger}f_{k} \label{Ham}
\end{eqnarray}

Here $i$ and $j$ are nearest neighbors and the hopping amplitude
$t_{ij}$ is $t$ along the $\hat y$ direction and alternating between
$t$ and $-t$ in the $\hat x$ direction from row to row; and $i$ and $k$ are second
nearest neighbors connected by hoppings along the square lattice diagonals, with amplitude
$\Delta_{ik} = i\Delta$ along the arrows and $\Delta_{ik} = -i\Delta$ against the arrows, see
Fig. 1. The unit cell contains two sublattices $A$ and $B$. This
model leads to a gapped state at half filling and the resulting
valence band has unit Chern number. This hopping model is equivalent
to a $d+id$ BCS state by an $SU\left(2\right)$ Gauge
transformation\cite{ludwig1994}. We use periodic boundary conditions
throughout this section.

\begin{figure}
\begin{centering}
\includegraphics[scale=0.35]{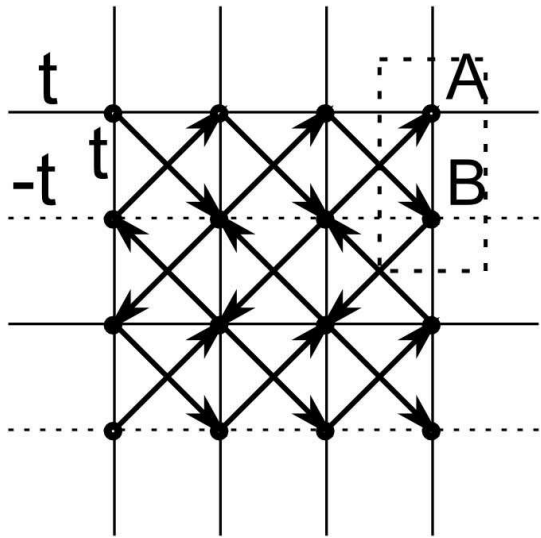}
\par\end{centering}
\caption{Illustration of a square lattice hopping model connected
with a $d+id$ superconductor. While the nearest neighbor hopping is
along the square edges with amplitude $t$ ($-t$ for hopping along
dashed lines), the second nearest neighbor hopping is along the square
diagonal (arrows in bold), with amplitude $+i\Delta$ ($-i\Delta$)
when hopping direction is along (against) the arrow. The two
sublattices in the unit cell are marked as $A$ and $B$.} \label{fig1}
\end{figure}

The unprojected ground state wave-function $|\phi\rangle$ is
obtained by filling all the states in the valence band ($\epsilon_k
< 0$) i.e. $|\phi \rangle =
\left[\underset{k,s}{\Pi}\gamma^\dagger_{sk}\right] |0\rangle$ where
$\gamma^\dagger_{k,s}=\psi_A(k)\underset{r_A}{\sum}f^\dagger_{r_A,s}e^{ik\cdot
r_A}+\psi_B(k)\underset{r_B}{\sum}f^\dagger_{r_B,s}e^{ik\cdot r_B}$
is the creation operator for a valence electron with spin $s$ and
momentum $k$, and $\psi_A(k)$($\psi_B(k)$) is the wave-function on
sublattice A(B). The projected wave-function that corresponds to the
chiral SL is obtained as $\left|\Phi\right\rangle =P|\phi \rangle$,
where $P$ is the Gutzwiller projection operator that projects the
wave-function to the Hilbert space of one electron per site. This is
implemented by restricting $\left|\alpha\right\rangle$ to the
Hilbert space of spins i.e. one particle per site. Due to the fact
that this Hamiltonian contains only real bipartite hoppings and
imaginary hoppings between the same sublattices and preserves the
particle-hole symmetry, this wave-function
$\left\langle\alpha|\Phi\right\rangle$ can be written as a product
of two Slater determinants $\mathcal M Det(M_{ij})^2$, where
$\mathcal M$ is just an unimportant Marshall sign factor, and:

\begin{eqnarray*}
M_{ij}=\left\{\left[\psi_A(k_i)+\psi_B(k_i)\right]+(-1)^{y_j}\left[\psi_A(k_i)-\psi_B(k_i)\right]\right\}e^{ik_i
\cdot r_j}
\end{eqnarray*}

$r_j$ is the coordinates of the up spins in configuration $\alpha$,
and $k_i$ is the momentums in the momentum space. The Renyi entropy
$S_{2}$ of this wave-function can be calculated by VMC method
detailed in the last section.

For an accurate calculation of TEE $\gamma$, it is important that
the subleading terms in the Eqn. \ref{eqntee0} be much smaller than
the universal constant $\gamma$ itself. This finite size error is
suppressed when the excitation gap is large and correlation length
is shorter than the system typical length scale. Note that the mean
field gap is given by $8\Delta$ for $\left|\Delta\right|\le 0.5t$
and $2t\sqrt{8-(t/\Delta)^2}$ for $\left|\Delta\right| > 0.5t$. To
minimize the finite size effect, we take $\Delta=0.5t$ unless
otherwise specified, so that the gap is large in both units of $t$
and $2\Delta$, and our calculation estimates a correlation length of
$\xi \sim 0.45$.

\subsection{Establishing Topological Order in chiral SL wave-function}

In this section we calculate the TEE $\gamma$ using the Kitaev-Preskill scheme \cite{kitaev2006}.

\begin{figure}
\begin{centering}
\includegraphics[scale=0.4]{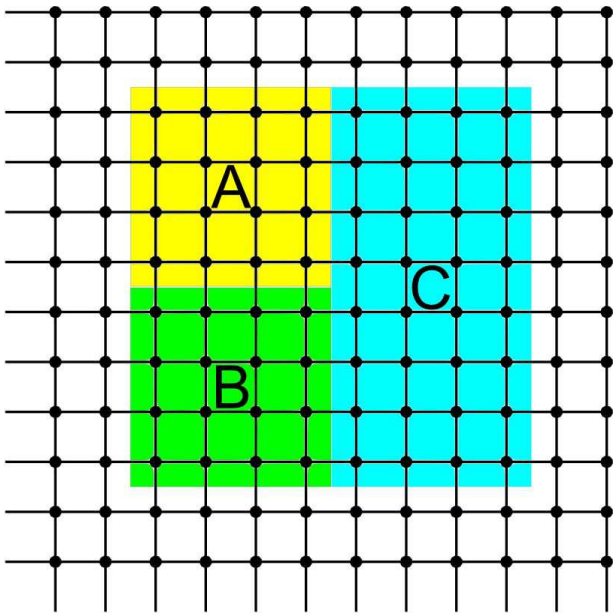}
\par\end{centering}
\caption{The separation of the system into subsystem $A$, $B$, $C$
and environment, periodic boundary condition is employed in both
$\hat{x}$ and $\hat{y}$ directions.} \label{fig3}
\end{figure}
We study system with total dimensions 12 $\times$ 12 lattice spacings
in both directions with periodic boundary conditions. We separate the system into $L_A\times L_A$ squares A
and B and an $L_A\times2L_A$ rectangle C, see Fig \ref{fig3}. For this particular geometry, TEE is simply given by:

\be
-\gamma = 2S_{2,A}-2S_{2,AC}+S_{2,ABC} \label{gamma1}
\ee

where we have used the fact that $S_{2,A}=S_{2,B}$,
$S_{2,AB}=S_{2,C}$ and $S_{2,AC}=S_{2,BC}$ owing to the  reflection
and translation symmetry of the wave function. This simplifies the
measurement of TEE into the measurement of $S_{2}$ for only three
subsystems $A$, $AC$ and $ABC$.

We use the unprojected wave-function as a benchmark for extraction
of TEE, which is non-interacting and hence exactly solvable. For an
$L_A=3$ system, the VMC calculation gives $\gamma=-0.0008\pm0.0059$,
in agreement with the absence of topological order and
correspondingly vanishing TEE(table \ref{table1} first row).

\begin{figure}
\begin{centering}
\includegraphics[scale=0.4]{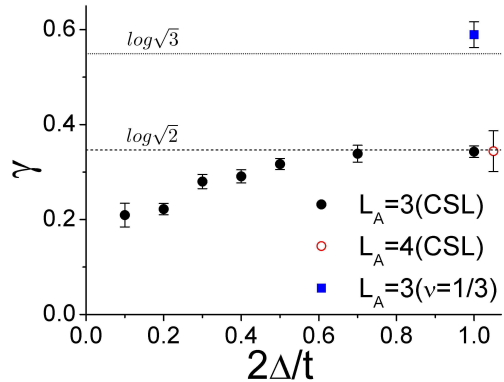}
\par\end{centering}
\caption{Illustration of finite size effect: chiral SL TEE $\gamma$
as a function of $2\Delta/t$ proportional to the relative gap size
for characteristic system length $L_A=3$. The larger the gap, the
closer the data approaches the ideal value. For comparison, TEE
$\gamma$ for chiral SL at $L_A=4$ and $2\Delta/t=1.0$ is shown. On
the same plot, TEE $\gamma$ of a lattice version of $\nu=1/3$
Laughlin state at $L_A=3$, $2\Delta/t=1.0$ is also shown. The dashed
lines are the ideal TEE values of $\gamma =
\log\left(\sqrt{2}\right)$ for the chiral SL and $\log\left(\sqrt{3}
\right)$ for the $\nu=1/3$ Laughlin state.} \label{fig6}
\end{figure}

The Gutzwiller projected wave-function is believed to be a chiral SL
which can be thought of as a Laughlin liquid at filling $\nu = 1/2$.
Using VMC method, we find $\gamma=0.343\pm0.012$ for an $L_A=3$
system and $\gamma=0.344\pm0.043$ for an $L_A=4$ system, both are in
excellent consistency with the expectation of
$\gamma=\log\left(\sqrt{2}\right)=0.347$ for its ground states' two
fold degeneracy, see table \ref{table1} second and third rows and
also Fig.\ref{fig6}.

We also want to point out that Gutzwiller projection qualitatively
changes the system ground state's topological and quantum behavior
from the mean field result.

On the other hand, by lowering the ratio of $2\Delta /t$ and
correspondingly the gap size the correlation length increases and
the finite size effects from subleading terms become more important.
See Fig.\ref{fig6} for the approach of the extracted TEE $\gamma$ to
its universal value of $\gamma =\log\left(\sqrt{2}\right)$ as we
lift the gap size controlled by $2\Delta/t$ for a system with
typical length scale $L_A=3$. The finite size analysis and the above
consistency between $L_A=3,4$ confirm that finite size effect is
small for our chosen sets of parameters for the system sizes we
study.

\section{A Lattice Version of the Laughlin State}
\label{sectionLaughlin}

Using VMC method, we further study the situations where the
wave-function is the cube or the fourth power of the Slater
determinant of the Chern insulator. For example, consider the
wave-function:

\begin{equation}\Psi_{1/3}(r_1,\,r_2\dots,\,r_N)=\Phi^3(r_1,\dots, \,r_N)\end{equation}

where $\Phi$ is the Chern insulator Slater determinant defined
above. Clearly, the product is a fermionic wave-function, since
exchanging a pair of particles leads to a sign change. This is
similar in spirit to constructing the corresponding Laughlin liquid
of $m=3$ of fermions, by taking the cube of the Slater determinant
wave-function in the lowest Landau level
$\psi(z_1,\dots,\,z_N)=\prod_{i<j}(z_i-z_j)e^{-\sum_i\frac{|z_i|^2}{4l^2_B}}$).
However, unlike the canonical Laughlin state, composed of lowest
Landau level states, these are rather different lattice
wave-functions. An interesting question is whether the lowest Landau
level structure is important in constructing states with the
topological order of the Laughlin state, or whether bands with
identical Chern number is sufficient, as suggested by field
theoretic arguments.

To address this we calculate TEE and compare with expectation for
the Laughlin phase. Again we choose $L_A=3$ in our VMC simulation,
and obtain: $\gamma=0.5894 \pm 0.0272$ for the $m=3$ wave-function,
in reasonable agreement with the ideal value
$\gamma=\log\left(\sqrt{3}\right)=0.549$(table \ref{table1} fourth
row).

We also considered the fourth power of the Chern insulator Slater determinant:
\begin{equation}\Psi_{1/4}(r_1,\,r_2\dots,\,r_N)=\Phi^4(r_1,\dots, \,r_N)\end{equation}

this is a bosonic wave-function, that is expected to be in the same
phase as $\nu=1/4$ bosons. Indeed we find with  $L_A=3$ in our VMC
simulation, $\gamma=0.732\pm0.076$, consistent with ideal value that
must be realized in the thermodynamic limit of this phase:
$\gamma=\log\left(\sqrt{4}\right)=0.693$(table \ref{table1} fifth
row).

These results offered direct support for the TEE formula
$\gamma=\log D$ as well as their validity as topological ground
state wave-functions carrying fractional charge and statistics. The
lattice fractional Quantum Hall wave-functions discussed here may be
relevant to the recent studies of flat band Hamiltonians with
fractional quantum Hall states \cite{tang2010,sheng2011}.

\section{Entanglement Entropy of a $Z_2$ spin liquid}
\label{Z2}

With the projected wave-function ansatz, we may also construct a
topological $Z_2$ SL by projecting another mean-field BCS state,
given by the specific BdG Hamiltonian on a square lattice as the
following\cite{wen2004}:

\begin{eqnarray*}
H=-\underset{\left\langle ij\right\rangle
}{\sum}\left(\psi_{i}^{\dagger}\mu_{ij}\psi_{j}+h.c.\right)+\underset{i}{\sum}\psi_{i}^{\dagger}a_{0}^{l}\tau^{l}\psi_{i}
\end{eqnarray*}

where
$\psi_{i}=\left(f_{\uparrow},f_{\downarrow}^{\dagger}\right)^{T}$.
$\tau^{1,2,3}$ are Pauli matrices. The second term is related to
chemical potentials, we set $a_{0}^{2,3}=0$, with $a_{0}^{1}$ fixed
by the conditions $\left\langle
\psi^{\dagger}\tau^{1,2,3}\psi\right\rangle =0$. Matrices $\mu_{ij}$
connecting nearest and next nearest neighbors:

\begin{eqnarray*}
\mu_{i,i+x}&=&\mu_{i,i+y}=-\tau^{3}\\
\mu_{i,i+x+y}&=&\eta\tau^{1}+\lambda\tau^{2}\\
\mu_{i,i-x+y}&=&\eta\tau^{1}-\lambda\tau^{2}
\end{eqnarray*}

This mean field model is readily solvable, with dispersion:

\begin{eqnarray*}
E_{k} &=& \sqrt{\epsilon_{k}^{2}+\left|\Delta_{k}^{2}\right|} \\
\epsilon_{k}&=&2\left(\cos\left(k_{x}\right)+\cos\left(k_{y}\right)\right)\\
\Delta_{k}&=&2\eta\left[\cos\left(k_{x}+k_{y}\right)+\cos\left(k_{x}-k_{y}\right)\right]+a_{0}^{1}\\
&
&-2i\lambda\left[\cos\left(k_{x}+k_{y}\right)-\cos\left(k_{x}-k_{y}\right)\right]
\end{eqnarray*}

We choose $\eta=\lambda=1.5$ for a large gap and our calculation
estimates that the correlation length is as short as $\xi \sim 1.3$
lattice spacings.

The VMC algorithm need little change\cite{gros1989}, except that
instead of Slater determinants product, the wave-function for spin
product configuration $\left|\alpha\right\rangle$ is given by

\begin{eqnarray*}
\phi_{\alpha}=\left\langle \alpha|\Phi\right\rangle
=\det\left(a_{ij}\right)
\end{eqnarray*}

here $a_{ij}=a\left(r_{i,\uparrow}-r_{j,\downarrow}\right)$ is the
Fourier transform of the superconducting pairing function$f_{k}$,
$r_{i,\uparrow}$ and $r_{j,\downarrow}$ are the coordinates of the
up-spins and down-spins, respectively:

\begin{eqnarray*}
f_{k}=\frac{\Delta_{k}}{\left|E_{k}+\epsilon_{k}\right|}
\end{eqnarray*}

For numerical simulations we again study $12\times12$ lattice
spacing systems and separate the system into subsystems including
$L_A\times L_A$ squares $A$ and $B$ and $L_A\times2L_A$ rectangle
$C$, again see Fig.\ref{fig3}. The TEE $\gamma$ is given by
Eq.\ref{gamma1} as before. First, we use the unprojected BCS state
as a benchmark, for which we expect a result of $\gamma=0.003$ from
an exact solution (since the unprojected state is a free particle
ground state, one may use the correlation matrix
method\cite{correlation-matrix}) and consistent with its absence of
topological order. Indeed, we obtain $\gamma=0.012\pm0.062$ using
VMC method, the almost vanishing value of $\gamma$ is consistent
with the expected value, which also serves as a check on our Monte
Carlo calculations.

On the other hand, the projection qualitatively alters the
topological properties of the system, and for simulation accuracy
and efficiency, we employ the 'sign trick' from
Ref.\cite{frank2011}. For an $L_A=4$ system, the VMC calculation
gives $\gamma = 0.584\pm0.089$. This is roughly consistent the $Z_2$
SL which has $D^2 =4$ sectors and $\gamma = \log\left(D\right)
=\log\left(2\right)\simeq 0.693$. The TEE is found to be about
$84\%$ of the expected value(table \ref{table1} last row). Other
studies on Z$_2$ phases eg. the quantum Monte Carlo on a Bose
Hubbard model in Ref.\cite{isakov2011} have also found values that
underestimate the topological entropy ($75\%$ of the expected zero
temperature value in that case). The smaller than expected value is
probably due to spinon excitations with a finite gap, causing
breaking of $Z_2$ electric field lines over the finite system size
$L_A=4$ we consider. Indeed, spin correlations decay more slowly for
the $Z_2$ state, as compared to the chiral SL, which also arrives
closer to its expected $\gamma$ value. Consistent with this fact is
the observation that for a {\em smaller} system size $L_A=3$ where
the finite system size has a larger impact, VMC calculation leads to
a value of $\gamma=0.446\pm0.119$, which is further away from the
ideal value.

\section{Conclusion}
In this paper we studied entanglement properties of candidate
wave-functions for $SU(2)$ symmetric gapped SLs and Laughlin states,
and established their topological order using the notion of TEE. We
studied two classes of SLs: 1) Wave-functions that describe quantum
Hall states and are obtained from the wave-function of a Chern
insulator by taking multiple copies of it 2) A $Z_2$ SL state that
is obtained by Gutzwiller projecting a fully-gapped BCS
superconductor. These wave-functions have long been used as ansatz
for exploring SLs states and it is reassuring that topologically
ordered states can be good variational ground states for realistic
Hamiltonians. Our method is directly applicable to any wave-function
that can be dealt within VMC method and would be especially useful
in cases where one is dealing with a Hamiltonian that has Monte
Carlo sign-problem and only has a variational ansatz for the
corresponding ground state. We also note that since the quantum Hall
wave-functions we study are not constructed from the lowest Landau
level but rather from the band structures that have non-zero Chern
number, our results are also relevant to the recently discovered
quantum Hall physics in flat band Hamiltonians \cite{tang2010,
sheng2011}.

Let us consider a few problems where our method may find immediate
application. Firstly, it would be interesting to apply our method to
$Z_2$ SLs that have gapless nodal spinons. These SLs are obtained by
Gutzwiller projecting a nodal BCS state. We note that in this case
one would find an additional contribution to the subleading constant
part of the entanglement entropy that comes from the gapless
spinons. Though we believe it might still be possible to separate
the total contribution of the constant term into a topological
constant and a term that comes from the gapless spinons only. It
would also be interesting to study wave-functions that are expected
to have non-abelian quasi-particles such as $SU(2)_k$ quantum Hall
wave-functions\cite{ronny1}. Thirdly, since VMC techniques can be
used for wave-functions defined in the continuum as well, it might
be interesting to study TEE of quantum Hall wave-functions (and
their descendants such as time-reversal invariant fractionalized
topological insulators \cite{levin2009} ) defined directly in the
continuum.

Finally, we note that one limitation of our method is that it can't
be used to calculate TEE for SLs where the gauge fields are coupled
to bosonic (rather than fermionic) spinons. This is because VMC
techniques are not very efficient when dealing with wave-functions
that are written as permanents (in contrast to determinants). It
would be interesting to see if the recent VMC calculation for a
$SU(2)$ symmetric bosonic SL \cite{tay2011} can be pushed to bigger
system sizes so as to establish topological order in such
wave-functions.

{\bf Acknowledgements:} We acknowledge support from NSF DMR- 0645691.


\begin{thebibliography}{7}
\bibitem{anderson1987} P.W. Anderson, Science 237, 1196 (1987).
\bibitem{kivelson1987} S. Kivelson, D. Rokhsar, J. Sethna, Phys. Rev. B 35,
8865 (1987).
\bibitem{read1989} N. Read, B. Chakraborty, Phys. Rev. B 40, 7133 (1989).
\bibitem{senthil2004} T. Senthil, A. Vishwanath, L. Balents, S. Sachdev and M. P. A. Fisher, Science 303, 1490 (2004).
\bibitem{exp} Y. Shimizu, K. Miyagawa, K. Kanoda, M. Maesato, and G. Saito, Phys. Rev. Lett. 91, 107001
(2003); Y. Okamoto, M. Nohara, H. Aruga-Katori and H. Takagi, Phys.
Rev. Lett. 99. 137207 (2007); J. S. Helton et al., Phys. Rev. Lett.
98, 107204 (2007); M. Yamashita et al, Science 328, 1246 (2010).
\bibitem{yan2010} Simeng Yan, David A. Huse, Steven R. White, Science 332, 1173 (2011).
\bibitem{meng2010} Z. Y. Meng, T. C. Lang, S. Wessel, F. F. Assaad, A. Muramatsu, Nature 464, 847 (2010).
\bibitem{wen2004} Xiao-Gang Wen, \textit{Quantum field theory of many-body systems}, Oxford Graduate Texts, 2004.
\bibitem{hamma2005} A. Hamma, R. Ionicioiu, and P. Zanardi, Phys. Lett. A 337, 22 (2005); Phys. Rev. A 71, 022315 (2005).
\bibitem{levin2006} M. Levin, X.-G. Wen, Phys. Rev. Lett. 96, 110405 (2006).
\bibitem{kitaev2006}  A. Kitaev, J. Preskill, Phys. Rev. Lett. 96, 110404 (2006).
\bibitem{rokhsar1988} D. S. Rokhsar and S. A. Kivelson, Phys. Rev. Lett. 61, 2376 (1988).
\bibitem{moessner2001} R. Moessner, S. L. Sondhi, Phys. Rev. Lett. 86, 1881 (2001).
\bibitem{senthil2002} T. Senthil, O. Motrunich, Phys. Rev. B 66, 205104 (2002).
\bibitem{balents2002} L. Balents, M.P.A. Fisher, S.M. Girvin, Phys. Rev. B 65, 224412 (2002).
\bibitem{misguich} G. Misguich, D. Serban and V. Pasquier, Phys. Rev. Lett. 89, 137202 (2002).
\bibitem{sachdev1992} N. Read and S. Sachdev, Phys. Rev. Lett. 66, 1773 (1991); S. Sachdev, Physical Review B 45, 12377 (1992).
\bibitem{Furukawa} S. Furukawa and G. Misguich, Phys. Rev. B 75, 214407 (2007).
\bibitem{Claudio} C. Castelnovo and C. Chamon, Phys. Rev. B 76, 174416 (2007).
\bibitem{Fradkin} S. Papanikolaou, K. S. Raman, and E. Fradkin, Phys. Rev. B 76, 224421 (2007).
\bibitem{hamma2008} A. Hamma, W. Zhang, S. Haas, and D. A. Lidar, Phys. Rev. B 77, 155111 (2008).
\bibitem{isakov2011} Sergei V. Isakov, Matthew B. Hastings, Roger G. Melko, arXiv:1102.1721.
\bibitem{varstudy} O. I. Motrunich, Phys. Rev. B 72, 045105 (2005); L. F. Tocchio, A. Parola, C. Gros, and F. Becca, Phys. Rev. B 80, 064419 (2009); T. Grover, N. Trivedi, T. Senthil and Patrick A. Lee, Phys. Rev. B 81, 245121 (2010).
\bibitem{kalmeyer} V. Kalmeyer and R. B. Laughlin, Phys. Rev. Lett. 59, 2095; V. Kalmeyer and R. B. Laughlin, Phys. Rev. B 39, 11 879; X. G. Wen, Frank Wilczek, and A. Zee, Phys. Rev. B 39,
11 413 (1989).
\bibitem{senthil2000} T. Senthil and Matthew P. A. Fisher, Phys. Rev. B 62, 7850 (2000).
\bibitem{wen1991} X.-G. Wen, Phys. Rev. B. 44, 2664 (1991).
\bibitem{frank2011} Yi Zhang, Tarun Grover, Ashvin Vishwanath, arXiv:1102.0350.
\bibitem{flammia2009} S. T. Flammia, A. Hamma, T. L. Hughes, and X.-G. Wen, Phys. Rev. Lett. 103, 261601 (2009).
\bibitem{dong} S. Dong, E. Fradkin, R. G. Leigh and S. Nowling, Journal of High Energy Physics 05 (2008) 016.
\bibitem{haque2007} M. Haque, O. Zozulya and K. Schoutens, Phys. Rev. Lett. 98, 060401 (2007); O. S. Zozulya, M. Haque, K. Schoutens, and E. H. Rezayi, Phys. Rev. B 76,
125310 (2007); A. M. Lauchli, E. J. Bergholtz and M. Haque, New J.
Phys. 12, 075004 (2010).
\bibitem{ivanov2002} D. A. Ivanov, T. Senthil, Phys. Rev. B 66, 115111 (2002). A. Paramekanti, M. Randeria and N. Trivedi, Phys. Rev. B 71, 094421 (2005).
\bibitem{Yao} Hong Yao and Xiao-Liang Qi, Phys. Rev. Lett. 105, 080501 (2010).
\bibitem{li2008} H. Li and F. D. M. Haldane, Phys. Rev. Lett. 101, 010504 (2008).
\bibitem{kitaev2003} A. Kitaev, Ann. Phys., 303, 2 (2003).
\bibitem{hastings2010} M. B. Hastings, I. Gonzalez, A. B. Kallin and R. G. Melko, Phys. Rev. Lett. 104, 157201 (2010).
\bibitem{gros1989} C. Gros, Annals of Physics 189, 53 (1989).
\bibitem{thomale} D. F. Schroeter, E. Kapit, R. Thomale, and M. Greiter, Phys. Rev. Lett. 99, 097202
(2007).
\bibitem{ludwig1994} A. W. W. Ludwig, M. P. A. Fisher, R. Shankar, G. Grinstein, Phys. Rev. B 50, 7526 (1994).
\bibitem{tang2010} E. Tang, J.-W. Mei, and X.-G. Wen, Phys. Rev. Lett. 106, 236802 (2011). T. Neupert, L. Santos, C. Chamon, and C. Mudry, Phys. Rev. Lett. 106, 236804 (2011); K. Sun, Z. C. Gu, H. Katsura, and S. Das Sarma, Phys. Rev. Lett. 106, 236803 (2011).
\bibitem{sheng2011} Y.-F. Wang, Z. C. Gu, C.-D. Gong, D. N. Sheng, arXiv:1103.1686; D. N. Sheng, Z. C. Gu, K. Sun, L. Sheng, Nature Communications 2, 389 (2011); N. Regnault, B. A. Bernevig, arXiv:1105.4867.
\bibitem{correlation-matrix} I. Peschel, V. Eisler, J. Phys. A: Math. Theor. 42, 504003 (2009).
\bibitem{ronny1} B. Scharfenberger, R. Thomale, M. Greiter, arXiv:1105.4348.
\bibitem{levin2009} M. Levin and A. Stern, Phys. Rev. Lett. 103, 196803 (2009).
\bibitem{tay2011} Tiamhock Tay, Olexei I. Motrunich, Phys. Rev. B 84, 020404(R) (2011).

\end{thebibliography}
\end{document}